\documentclass[journal]{IEEEtran}
\usepackage{amsthm}
\usepackage{amsfonts}
\usepackage{amssymb}
\ifCLASSINFOpdf
  \usepackage[pdftex]{graphicx}
  \usepackage{graphicx}
  \graphicspath{{../pdf/}{../jpeg/}{../png/}}
  \DeclareGraphicsExtensions{.pdf,.jpeg,.png}
\else
  \usepackage[dvips]{graphicx}
  \graphicspath{{../eps/}}
  \DeclareGraphicsExtensions{.eps}
\fi
\ifCLASSOPTIONcompsoc
  \usepackage[caption=false,font=normalsize,labelfont=sf,textfont=sf]{subfig}
\else
\usepackage[caption=false,font=footnotesize]{subfig}
\fi
\usepackage{multirow}
\usepackage{balance}
\usepackage{multicol}
\usepackage{epstopdf}
\usepackage{verbatim} 

\usepackage[english]{babel}
\usepackage{float}
\usepackage{xcolor}
\usepackage[linesnumbered,ruled,vlined]{algorithm2e}

\usepackage[hidelinks]{hyperref}
\usepackage{cite}
\ifCLASSINFOpdf
   \usepackage[pdftex]{graphicx}
   \else
\usepackage[dvips]{graphicx}
\fi
\usepackage[cmex10]{amsmath}
\usepackage[T1]{fontenc}
\newcommand*{\affmark}[1][*]{\textsuperscript{\dag}}

\begin{document}
\begin{center}
    {\normalsize \textbf{Two-Stage Hybrid Transceiver Design Relying on Low-Resolution ADCs in Partially Connected MU Terahertz (THz) MIMO Systems}}\vspace{-0.1\baselineskip}
\end{center}
\begin{center}
    Abhisha Garg, \textit{Graduate Student Member, IEEE}, Akash Kumar, Suraj Srivastava, \textit{Member, IEEE}, Aditya K. Jagannatham, \textit{Senior Member, IEEE}, Lajos Hanzo, \textit{Life Fellow, IEEE}\vspace{-0.1\baselineskip}
\end{center}
\footnote{The work is supported by IEEE SPS scholarship grant for $2023, 2024$ and $2025$. The work of Aditya K. Jagannatham was supported in part by the Qualcomm Innovation Fellowship; in part by the Qualcomm 6G UR Gift; in part by the Arun Kumar Chair Professorship. The work of S. Srivastava was supported in part by Anusandhan National Research Foundation’s PM-ECRG/2024/478/ENS; and in part by Telecom Technology Development Fund (TTDF) under Grant TTDF/6G/368. S. Srivastava and A. K. Jagannatham jointly acknowledge the funding support provided by Anusandhan National Research Foundation's Advanced Research Grant ANRF/ARG/2025/005895/ENS. The work of Lajos Hanzo is supported by Engineering and Physical Sciences Research Council (EPSRC) projects is gratefully acknowledged: Platform for Driving Ultimate Connectivity (TITAN) (EP/X04047X/1; EP/Y037243/1); Robust and Reliable Quantum Computing (RoaRQ, EP/W032635/1); PerCom (EP/X012301/1); India-UK Intelligent Spectrum Innovation ICON UKRI-1859. S. Srivastava, A. K. Jagannatham and Lajos Hanzo jointly acknowledge the funding support provided to ICON-project by DST and UKRI-EPSRC under India-UK Joint opportunity in Telecommunications Research.

Abhisha Garg and Aditya K. Jagannatham are with the Department of Electrical Engineering, Indian Institute of Technology Kanpur, Kanpur-$208016$, India (e-mail: abhisha20@iitk.ac.in; adityaj@iitk.ac.in). 

Akash Kumar is with Qualcomm India Pvt. Ltd., Hyderabad, Telangana, $500081$, India (email: akkum@qti.qualcomm.com). 

Suraj Srivastava is with the Department of Electrical Engineering, Indian Institute of Technology Jodhpur, Jodhpur, Rajasthan $342030$, India (e-mail: surajsri@iitj.ac.in). 

L. Hanzo is with the School of Electronics and Computer Science, University of Southampton, Southampton SO17 1BJ, U.K. (email: lh@ecs.soton.ac.uk)}
\begin{abstract}
A two-stage hybrid transceiver is designed by considering a partially connected architecture at the base station (BS) for a low-resolution multi-user (MU) THz massive multiple input multiple output (MIMO) system. Due to its high bandwidth coupled with a high number of antennas, the THz band suffers from the deleterious \textit{spatial-wideband} and \textit{frequency-wideband effects} jointly termed as the \textit{dual-wideband effect.} To address this undesired phenomenon, we rigorously model the THz MIMO channel at each subarray corresponding to each user by incorporating the absorption, reflection, and free-space losses. Subsequently, a novel beamforming technique is proposed that employs only a few true time delay (TTD) lines for eliminating the \textit{beam-split effect}, which is the manifestation of the spatial-wideband effect in the frequency domain. Our simulation results demonstrate a performance improvement of around $13\%$ in terms of spectral efficiency over the existing state-of-the-art techniques. 
\end{abstract}
\vspace{-4.5mm}
\IEEEpeerreviewmaketitle
\vspace{-6.5mm}
\section{Introduction} \label{intro}
\vspace{-1mm}
Terahertz (THz) communication spanning the frequency range of $0.3-10$ THz \cite{garg2024angularly}, is poised to revolutionize next-generation wireless communication by offering a large bandwidth and extreme data rates. A typical wideband THz channel is frequency-selective in nature, due to the multipath-induced delay spread, which is a characteristic feature of wireless signal propagation. This phenomenon is termed as the frequency-wideband effect. Due to the short wavelength of the THz signal, a large antenna array leads to a progressively increasing phase shift in the received/ transmitted signal across the constituent antenna elements, corresponding to the propagation distance in the time domain, termed the spatial-wideband effect \cite{gao2021wideband}. Collectively, these detrimental effects are referred to as the dual-wideband effect. Both the frequency-wideband and the spatial-wideband effects independently contribute to performance degradation in the THz systems. The spatial-wideband effect experienced in the frequency domain leads to a variation in angle of arrival/ departure (AoA/ AoD) across the subcarrier frequency range, which results in the \textit{beam-split} effect \cite{dai2022delay}. Zhang \textit{et al.} \cite{zhang2021analysis} compare the performance of finite-bit ADCs and DACs in the context of narrow-band, partially connected MU THz hybrid MIMO systems. Forsch \textit{et al.} \cite{forsch2025detection} proposed novel optimal and suboptimal detection schemes for THz band systems with uniform multi-bit quantization in their seminal work. Nikbakht and Lozano, associated \cite{nikbakht2021terahertz} proposed an unsupervised learning-based technique for transmit beamforming in THz systems considering low-resolution ADCs. However, none of the existing works on low-resolution ADCs consider the dual-wideband effect that is predominant in the THz band. In their groundbreaking work \cite{gao2021wideband}, the authors proposed an effective beamforming technique based on TTD designed for THz MIMO-OFDM systems to compensate for the beam-split effect. While the existing literature successfully addresses certain aspects of beam-split compensation, some critical challenges remain unresolved with respect to practical implementation. These include the incorporation of low-resolution analog-to-digital converters (ADCs) and the design as well as implementation of partially connected hybrid architectures, which further exacerbate the beam-split effect. The next section summarizes the key contributions of this compact letter.
\begin{table*}[hbt!] 
    \centering
      \caption{Boldly contrasting our contributions to the literature}\vspace{-0.8\baselineskip}
\begin{tabular}{|l|c|c|c|c|c|c|c|c|c|c|c|c|c|c|c|c|}    \hline

\textbf{Features} &\cite{zhang2021analysis} & \cite{forsch2025detection} &\cite{nikbakht2021terahertz} & \cite{choi2019two}& \cite{gao2024energy} & \cite{roth2017achievable} &\cite{roth2018comparison} & \cite{gao2021wideband} &\textbf{This paper} \\ 

 \hline

THz hybrid MIMO

& \checkmark & \checkmark & \checkmark &  &  &  & & \checkmark & \checkmark \\

 \hline

mmWave hybrid MIMO

&  &  & & \checkmark & \checkmark & \checkmark & \checkmark & & \\

 \hline
 
MU-MIMO

& \checkmark &  & &   & \checkmark &  & \checkmark & & \checkmark\\

 \hline
 
Partially connected architecture

& \checkmark &  &  &  & \checkmark & \checkmark & \checkmark & & \checkmark \\
\hline

Bussgang Decomposition

& \checkmark &  &  & \checkmark & \checkmark & \checkmark & \checkmark & & \checkmark \\
 \hline

Dual-Wideband Effect with TTD lines

& &  & &  &  &  & & \checkmark & \checkmark \\

 \hline

SC-FDE system with low-resolution ADCs

&  &  &  &  & &  & &  & \checkmark \\
 \hline

Partially connected with Bussgang decomposition and dual-wideband 

&  &  &  &  & &  & &  & \checkmark \\
 \hline
    \end{tabular}\vspace{-2\baselineskip}
     \label{con}
\end{table*}
\vspace{-3.5mm}
\subsection{Contributions} \label{contri}
\vspace{-1mm}
\begin{itemize}
    \item Our work considers a partially connected MU-MIMO architecture employing low-resolution ADCs at the BS. Furthermore, the spatial wideband effect is considered, which is a characteristic feature of THz signals, together with the frequency wideband effect, and the Root Raised Cosine pulse shaping filter (RRC-PSF). To the best of our knowledge, none of the papers in the existing literature consider an single-carrier frequency domain equalization (SC-FDE)-based system with low-resolution ADCs in the context of dual-wideband channels, wherein SC systems are preferred over their OFDM counterparts \cite{mollen2016one}. Additionally, the proposed framework employs the well-established Bussgang decomposition for the linearization of the nonlinear quantized system model, which is crucial for approaching capacity. Furthermore, it is worth noting that there is no existing study of a practical wideband precoding solution that considers the partially connected architecture with Bussgang decomposition-based linearization in the face of the dual-wideband effects that are endemic to THz systems.
    \item Additionally, a novel two-stage hybrid beamforming scheme is developed for the proposed partially connected low-resolution THz MU-MIMO system. In Stage-1, the optimal beam-steering angles corresponding to each RF chain are computed using the spatially sparse algorithm, which exploits the angular sparsity of the THz MIMO channel. The algorithm selects the optimal beamforming vector from the dictionary matrix corresponding to each subarray. Stage-2 subsequently uses these angles to determine the appropriate phase shifts required for transmission at each RF chain using the TTD elements, which convert the traditional frequency-invariant PSs to their frequency-dependent equivalents. Therefore, via aligning the optimal beamforming direction with the physical ray direction, the beam-split effect can be substantially mitigated by harnessing the time delay elements.
    \item The performance of the proposed framework is extensively compared in terms of the spectral efficiency attained for RRC-PSF and rectangular pulse shaping filter (Rect-PSF) based dual-wideband THz channels, leading to the observation of a trade-off between both these channel formulations. It is evident that the proposed approach offers a significant performance improvement over the existing techniques, and substantially mitigates the beam-split effect. In Table-I we boldly contrast our contributions to the literature.
\end{itemize}
\vspace{-3.5mm}
\subsection{Notation} 
\vspace{-1mm}
Matrices are represented by uppercase letters $\mathbf{B}$, while lowercase letters $\mathbf{b}$ represent vectors. Various operators are indicated by superscripts such as $(.)^T, (.)^H, (.)^{-1}$ which represent the transpose, Hermitian and inverse, respectively. The term $\mathrm{blkdiag}(\mathbf{B}_1, \mathbf{B}_2, \cdots, \mathbf{B}_N)$ describes a block-diagonal matrix associated with $\mathbf{B}_1, \mathbf{B}_2, \cdots, \mathbf{B}_N$ on the principal diagonal. The quantities $\mathcal{C}(.)$ and $\mathcal{R}(.)$ denote the column and row spaces of a matrix. The operator $\mathcal{CN}(\boldsymbol{\mu},\mathbf{R})$ represents the complex Gaussian distribution with mean $\boldsymbol{\mu}$ and covariance $\mathbf{R}$. Let $\left\{\mathbf{H}_{u}(0), \mathbf{H}_{u}(1), \cdots, \mathbf{H}_{u}(N-1)\right\}$ represent a sequence of matrices and $\left\{\mathbf{x}_u(0), \mathbf{x}_u(1), \cdots, \mathbf{x}_u(N-1)\right\}$ represent a sequence of vectors. The circular convolution $\left\{\mathbf{r}_u(n)\right\}_{n=0}^{N-1}$ can be defined as
\vspace{-4mm}
    \begin{equation}
        \begin{aligned}
        \mathbf{r}_u(n) & = \sum_{l=0}^{N-1} \mathbf{H}_{u}(l)\mathbf{x}_u[(n-l)]_N + \check{\mathbf{v}}_u(n),
    \end{aligned}
    \end{equation}
    where $[.]_N$ represents modulo-N operation. Let $\left\{\mathbf{x}_u(0), \mathbf{x}_u(1), \cdots, \mathbf{x}_u(N-1)\right\}$ represent the input sequence, where $\mathbf{x}(n,k)$ denotes the $k$th element of $\mathbf{x}(n)$. Let $\left\{\mathbf{r}_u(0), \mathbf{r}_u(1), \cdots, \mathbf{r}_u(N-1)\right\}$ constitute the output sequence, where $\mathbf{r}(n,k)$ denotes the $k$th element of $\mathbf{r}(n)$. Then, the N-point FFT of the vector sequence is defined as $\mathbf{r}(p,q) = \sum_{n=0}^{N-1}\mathbf{x}(n,q)e^{-j\frac{2 \pi n p}{N}}$.
\vspace{-1mm}
\section{Partially connected MU THz MIMO system and channel models} \label{parsys}
Consider a SC-FDE aided \cite{srivastava2021fast} wideband THz MU MIMO system, where the BS is equipped with $N_{\mathrm{sub}} = N_{\mathrm{RF}}^R$ subarrays, each consisting of $N_{\mathrm{BS}}^{s_l}$ antennas. Explicitly, we have $s_l = 1, 2, \cdots, N_{\mathrm{RF}}^R$, where $N_{\mathrm{RF}}^R$ denotes the number of RF chains at the BS. Therefore, the total number of antennas at the BS is $N_{\mathrm{BS}} = \sum_{s_l=1}^{N_{\mathrm{RF}}^R} N_{\mathrm{BS}}^{s_l}$. Furthermore, let the BS simultaneously serve $U$ users, with each user transmitting $N_{s,u}$ data streams. The total number of data streams at the BS is $N_s = \sum_{u=1}^U N_{s,u}$. Additionally, the $u$-th user possesses $N_{T,u}$ antennas in a fully-connected architecture along with $N_{\mathrm{RF},u}^T$ RF chains. Therefore, the number of transmit antennas is $N_T$ across all users is $N_T = \sum_{u=1}^U N_{T,u}$ and the number of transmit RF chains is $N_{\mathrm{RF}}^T = \sum_{u=1}^U N_{\mathrm{RF},u}^T$. Let $\mathbf{H}_{l,u}^{s_l} \in \mathbb{C}^{N_{\mathrm{BS}}^{s_l} \times N_{T,u}}$ represent the complex wideband THz channel corresponding to the $l$-th delay tap of the $u$-th user on the $s_l$-th subarray. Furthermore, let $\mathbf{b}_{u}^{(d)} \in \mathbb{C}^{N_{s,u} \times 1}$, $0 \leq d \leq N_d - 1$, represent the $d$-th complex data vector corresponding to the $u$-th user. Prior to transmission, the data vectors undergo zero-padding (ZP), where $L-1$ zeros are appended to each block to generate a ZP block of length $K = N_d+L-1$, which is given by $\Big\{\mathbf{b}_{u}^{(a)} \Big\}_{a=0}^{K-1} = \Big\{\mathbf{b}_{u}^{(0)}, \mathbf{b}_{u}^{(1)}, \cdots, \mathbf{b}_{u}^{(N_d-1)}, \underbrace{\mathbf{0},\cdots,\mathbf{0}}_{L-1} \Big\}$. Furthermore, $N_d-1$ zero-matrices of size $N_{\mathrm{BS}}^{s_l} \times N_{T,u}$ are appended to the channel taps $\mathbf{H}_{l,u}^{s_l}$, which are given by $\left\{\mathbf{H}_{a,u}^{s_l}\right\}_{a=0}^{K-1} = \Big\{\mathbf{H}_{0,u}^{s_l}, \mathbf{H}_{1,u}^{s_l}, \cdots, \mathbf{H}_{L-1,u}^{s_l},\underbrace{\mathbf{0},\cdots,\mathbf{0}}_{N_d - 1} \Big\}$. Additionally, the overall channel matrix $\mathbf{H}_{a,u} \in \mathbb{C}^{N_{\mathrm{BS}} \times N_{T,u}}$ corresponding to the $u$-th user at the BS is defined as $\mathbf{H}_{a,u} = \big[\left(\mathbf{H}_{a,u}^1\right)^T \, \big(\mathbf{H}_{a,u}^2\big)^T \, \cdots \, \big(\mathbf{H}_{a,u}^{N_{\mathrm{RF}}^R}\big)^T\big]^T$, where $\mathbf{H}_{a,u}^{s_l} \in \mathbb{C}^{N_{\mathrm{BS} }^{s_l}\times N_{T,u}}$ denotes the channel between the $s_l$-th subarray and the $u$-th user for the $a$-th time instant. Therefore, the received signal vector $\mathbf{r}(a) \in \mathbb{C}^{N_{\mathrm{BS}} \times 1}$ corresponding to all the $U$ users is given as
\vspace{-2mm}
\begin{equation}
    \mathbf{r}(a) = \sum_{u=1}^U \mathbf{H}_{a,u} \otimes_K \left(\mathbf{F}_{\mathrm{RF},u}\mathbf{F}_{\mathrm{BB},u} \mathbf{b}_{u}^{(a)} \right) + \check{\mathbf{v}}(a),
\end{equation}
where $\mathbf{H}_{a,u}$ represents the sequence of matrices, while the quantity $\mathbf{F}_{\mathrm{RF},u}\mathbf{F}_{\mathrm{BB},u} \mathbf{b}_{u}^{(a)}$ represents the sequence of vectors corresponding to the $u$-th user. The quantity $\check{\mathbf{v}}(a)$ represents the AWGN noise that obeys the distribution $\mathcal{CN}\left(\mathbf{0}_{N_{\mathrm{BS}} \times 1}, \sigma^2 \mathbf{I}_{N_{\mathrm{BS}}}\right)$ and $\otimes_K$ represents the circular convolution of length $K$. The quantity $\mathbf{F}_{\mathrm{RF},u} \in \mathbb{C}^{N_{T,u} \times N_{\mathrm{RF},u}^T}$ and $\mathbf{F}_{\mathrm{BB},u} \in \mathbb{C}^{N_{\mathrm{RF},u}^T \times N_{s,u}}$ represent the RF and baseband transmit precoders (TPC) corresponding to the $u$-th user. Note that the ZP operation converts the linear convolution to circular convolution, and by further applying the fast Fourier transform (FFT) operation at the receiver, the system is decoupled into equivalent frequency bins \cite{garg2024angularly}. It is worth noting that although the overall system model is decoupled into different frequency bins at the receiver, interestingly, it can be viewed as a single carrier system at the transmitter. This is due to the fact that there is no inverse FFT (IFFT) block at the transmitter, and the transmit symbols $\mathbf{b}_u^{(d)}$ are conveyed serially with ZP. Furthermore, the received signal $\mathbf{y}(a) \in \mathbb{C}^{N_{\mathrm{RF}}^R \times 1}$ of all the users after performing RF combining with the aid of $\mathbf{W}_{\mathrm{RF}}^H \in \mathbb{C}^{N_{\mathrm{BS}} \times N_{\mathrm{RF}}^R}$ and passing through low-resolution ADCs $\mathcal{Q}(.)$ is given as
\vspace{-3mm}
\begin{equation}
    \begin{aligned}
    \mathbf{y}(a) = \mathcal{Q}\Big(\mathbf{W}_{\mathrm{RF}}^H \sum_{u=1}^U \mathbf{H}_{a,u} \otimes_k \big(\mathbf{F}_{\mathrm{RF},u} \mathbf{F}_{\mathrm{BB},u}\mathbf{b}_u^{(a)} \big)+ \mathbf{W}_{\mathrm{RF}}^H\check{\mathbf{v}}(a)\Big), \nonumber
\end{aligned}
\end{equation}
\vspace{-8mm}
\begin{align}
    \approx \mathbf{A}\mathbf{W}_{\mathrm{RF}}^H  \sum_{u=1}^U \mathbf{H}_{a,u} \otimes_k \big(\mathbf{F}_{\mathrm{RF},u} \mathbf{F}_{\mathrm{BB},u}\mathbf{b}_u^{(a)} \big)+ \mathbf{A}\mathbf{W}_{\mathrm{RF}}^H\check{\mathbf{v}}(a) + \tilde{\mathbf{v}}_q. \nonumber
\end{align}
The subsequent approximation is due to the widely used Bussgang decomposition \cite{mezghani2012capacity}, which models the non-linear quantized output by a resolution of $b$-bits. Note that, for a partially-connected architecture at the BS, the analog RF receive combiner (RC) matrix $\mathbf{W}_{\mathrm{RF}}$ is formulated as $\mathbf{W}_{\mathrm{RF}} = \mathrm{blkdiag}\big(\mathbf{w}_{\mathrm{RF}}^1 \mathbf{w}_{\mathrm{RF}}^2 \cdots \mathbf{w}_{\mathrm{RF}}^{N_{\mathrm{RF}}^R}\big)$, where $\mathbf{w}_{\mathrm{RF}}^{s_l} \in \mathbb{C}^{N_{\mathrm{BS}}^{s_l} \times 1}$ represents the analog RF RC vector associated with the $s_l$-th RF chain. Moreover, the elements of the RF TPC and RC follow the constant-magnitude constraint $\left|\mathbf{F}_{\mathrm{RF},u} (\rho,\kappa)\right| = \frac{1}{\sqrt{N_{T,u}}}, \left|\mathbf{W}_{\mathrm{RF}}(\rho,\kappa)\right| = \frac{1}{\sqrt{N_{\mathrm{BS}}^{s_l}}} \: \forall \: \rho,\kappa$. The operator $\mathcal{Q}(.)$ represents the quantization operation. The matrix $\mathbf{A} \in \mathbb{C}^{N_{\mathrm{RF}}^R \times N_{\mathrm{RF}}^R}$ is assumed to be diagonal under the assumption that each RF chain is quantized independently. Its elements are defined as $\mathbf{A} = \xi \mathbf{I}$, where $\xi = {1-\Tilde{\rho}}$ and $\Tilde{\rho}$ denotes the noise-to-signal ratio, as given in \cite{fan2015uplink} (Table I). Let $\mathbb{E}(\mathbf{b}^{(a)}_u(\mathbf{b}^{(a)}_u)^H) = \sigma_b^2 \mathbf{I}_{N_s}$. The additive quantization noise $\Tilde{\mathbf{v}}_q$ is modeled as a complex Gaussian random vector following the distribution $\mathcal{CN}(\mathbf{0}_{N_{\mathrm{RF}}^R \times 1}, \mathbf{R})$, where the diagonal covariance matrix $\mathbf{R}$ is given by $\mathbf{R} = \xi(1-\xi) \mathrm{diag}(\mathbf{W}_{\mathrm{RF}}^H \Tilde{\mathbf{Q}} \mathbf{W}_{\mathrm{RF}} + \sigma_n^2 \mathbf{W}_{\mathrm{RF}}^H \mathbf{W}_{\mathrm{RF}})$ and $\Tilde{\mathbf{Q}} = \sum_{u=1}^U\sum_{n=0}^{K-1} \mathbf{H}_u(n)\mathbf{R}_{uu}\mathbf{H}_u^H(n)$ with $\mathbf{R}_{uu} = \sigma_b^2 \mathbf{F}_{\mathrm{RF},u} \mathbf{F}_{\mathrm{BB},u} \mathbf{F}_{\mathrm{BB},u}^H \mathbf{F}_{\mathrm{RF},u}^H$. Let $\boldsymbol{\eta}(a) = \mathbf{A}\mathbf{W}_{\mathrm{RF}}^H\check{\mathbf{v}}(a) + \Tilde{\mathbf{v}}_q$ denote the combined effective noise, which follows the distribution $\mathcal{N}(\mathbf{0}_{N_{\mathrm{RF}}^R \times 1}, \mathbf{C})$ where $\mathbf{C} = \xi^2 \sigma_n^2 \mathbf{W}_{\mathrm{RF}}^H \mathbf{W}_{\mathrm{RF}} + \xi(1-\xi) \mathrm{diag}(\mathbf{W}_{\mathrm{RF}}^H \Tilde{\mathbf{Q}} \mathbf{W}_{\mathrm{RF}} + \sigma_n^2 \mathbf{W}_{\mathrm{RF}}^H \mathbf{W}_{\mathrm{RF}})$. Note that, the diagonal error covariance matrix results from quantizing each RF chain independently. The off-diagonal elements in the covariance matrix are zero, indicating that the quantization errors across different RF chains are uncorrelated. Also, we assume the received power per RF chain is statistically identical. As a result, the effective noise covariance is approximated as scaled identity matrix. This follows from the properties of the Bussgang decomposition for Gaussian inputs \cite{mezghani2012capacity}. Therefore, the received signal $\mathbf{y}[k] \in \mathbb{C}^{N_{\mathrm{RF}}^R \times 1}$ corresponding to the $k$-th frequency bin for all the users, obtained via the $K$-point FFT i.e., $ \big\{\mathbf{y}[k]\big\}_{k=0}^{K-1} = \mathrm{FFT}\big(\big\{\mathbf{y}(a)\big\}_{a=0}^{K-1}\big)$, is determined as
 \vspace{-3mm}
\begin{align}
    \mathbf{y}[k] \approx \mathbf{A} \mathbf{W}_{\mathrm{RF}}^H \sum_{u=1}^U \mathbf{H}_u[k] \mathbf{F}_{\mathrm{RF},u} \mathbf{F}_{\mathrm{BB},u} \mathbf{b}_{u}[k] + \boldsymbol{\eta}[k].
\end{align}
The quantity $\mathbf{b}_{u}[k] \in \mathbb{C}^{N_{s,u} \times 1}$ represents the equivalent transmit signal vector corresponding to the $u$-th user at the $k$-th frequency bin, which is the $k$-th output of the $K$-point FFT of the time domain signal block $\big\{\mathbf{b}_u^{(a)}\big\}_{a=0}^{K-1}$. Moreover, the quantity $\boldsymbol{\eta}[k] \in \mathbb{C}^{N_{\mathrm{RF}}^R \times 1}$ represents the equivalent noise power, which is the $K$-point FFT of $\big\{\boldsymbol{\eta}^{(a)}\big\}_{a=0}^{K-1}$. The overall channel matrix $\mathbf{H}_u[k] \in \mathbb{C}^{N_{\mathrm{BS}} \times N_{T,u}}$ represents the $K$-point FFT of $\{\mathbf{H}_{l,u}\}_{l=0}^{K-1}$. The signal $\mathbf{y}[k] \in \mathbb{C}^{N_s \times 1}$ received at the BS after baseband combining with $\mathbf{W}_{\mathrm{BB}}[k] \in \mathbb{C}^{N_{RF}^R \times N_s}$ is given by
\vspace{-1.5mm}
\begin{equation}
    \begin{aligned}
    \mathbf{y}[k] \approx \mathbf{W}_{\mathrm{BB}}^H[k] & \mathbf{A}\mathbf{W}_{\mathrm{RF}}^H \mathbf{H}_{\mathrm{MU}}[k] \, \tilde{\mathbf{F}}_{\mathrm{RF}} \tilde{\mathbf{F}}_{\mathrm{BB}} \tilde{\mathbf{b}}[k]  + \mathbf{W}_{\mathrm{BB}}^H[k]\boldsymbol{\eta}[k],
\end{aligned}
\end{equation}
where $\mathbf{H}_{\mathrm{MU}}[k] = \left[\mathbf{H}_1[k] \; \mathbf{H}_2[k] \; \cdots \; \mathbf{H}_U[k]\right] \in \mathbb{C}^{N_{\mathrm{BS}} \times N_T}$ represents the concatenated channel across all the users. Furthermore, $\tilde{\mathbf{F}}_{\mathrm{RF}} = \mathrm{blkdiag}\left(\mathbf{F}_{\mathrm{RF},1}, \mathbf{F}_{\mathrm{RF},2},\cdots,\mathbf{F}_{\mathrm{RF},U}\right) \in \mathbb{C}^{N_T \times N_{\mathrm{RF}}^T}$ denotes the concatenated RF TPC, while $\tilde{\mathbf{F}}_{\mathrm{BB}} = \mathrm{blkdiag}\left(\mathbf{F}_{\mathrm{BB},1}, \mathbf{F}_{\mathrm{BB},2}, \cdots, \mathbf{F}_{\mathrm{BB},U}\right) \in \mathbb{C}^{N_{\mathrm{RF}}^T \times N_s}$ represents the compound baseband TPC designed for all the users and $\tilde{\mathbf{b}}[k] = \left[\mathbf{b}_1^T[k] \; \mathbf{b}_2^T[k] \; \cdots \; \mathbf{b}_U^T[k]\right]^T \in \mathbb{C}^{N_s \times 1}$ is the $N$-point FFT of the concatenated pilot vector at the $k$-th frequency bin. Let $\tilde{\boldsymbol{\eta}}[k] = \mathbf{W}_{\mathrm{BB}}^H[k]\boldsymbol{\eta}[k]$ represent the combined effective noise, which obeys the distribution $\mathcal{CN}(\mathbf{0}_{N_s \times 1}, \tilde{\mathbf{C}}[k])$, where $\tilde{\mathbf{C}}[k]$ is given as $\tilde{\mathbf{C}}[k] = \mathbf{W}_{\mathrm{BB}}^H[k] \, \mathbf{C} \, \mathbf{W}_{\mathrm{BB}}[k]$. Thus, one can define the sum spectral efficiency corresponding to all the users as
\vspace{-5mm}
\begin{align}
    \mathrm{SE} = \frac{1}{K} \sum_{k=0}^{K-1} \mathrm{log}_2 \Big|\mathbf{I}_{N_s} + \frac{1}{N_s} \tilde{\mathbf{C}}^{-1}[k] \mathbf{S}[k] \Big|, \label{Spec_eff}
\end{align}
where $\mathbf{S}[k] = \mathbf{W}^H[k] \mathbf{H}_{\mathrm{MU}}[k] \mathbf{F}\mathbf{F}^H \mathbf{H}^H_{\mathrm{MU}}[k]\mathbf{W}[k]$, $\mathbf{W}[k] = \mathbf{W}_{\mathrm{RF}} \mathbf{A}^H \mathbf{W}_{\mathrm{BB}}[k]$ and $\mathbf{F} = \tilde{\mathbf{F}}_{\mathrm{RF}} \tilde{\mathbf{F}}_{\mathrm{BB}}$. The next section describes dual-wideband THz channel.
\vspace{-4mm}
\subsection{Dual-wideband THz channel model}
\vspace{-1mm}
The array response vector $\tilde{\mathbf{a}}_{\mathrm{BS}}^{s_l}(\tilde{\theta}^{s_l}, f_k)$ constructed for subarray $s_l$ with $N_{\mathrm{BS}}^{s_l}$ elements, after incorporating the spatial-wideband effect, is determined as
\vspace{-2mm}
\begin{align}
    \tilde{\mathbf{a}}_{\mathrm{BS}}^{s_l}\big(\tilde{\theta}^{s_l},f_k\big) = \frac{1}{\sqrt{N_{\mathrm{BS}}^{s_l}}} & \big[1,e^{-j \pi \frac{f_k}{f_c} \sin{\tilde{\theta}^{s_l}}},e^{-j \pi 2 \frac{f_k}{f_c} \sin{\tilde{\theta}^{s_l}}}, \notag \\ & \cdots, e^{-j \pi (N_{\mathrm{BS}}^{s_l}-1) \frac{f_k}{f_c} \sin{\tilde{\theta}^{s_l}}} \big]^T, \label{arra_res}
\end{align}
where $f_k = f_c + \big(k-\frac{K+1}{2}\big)\frac{B}{K}$ represents the subcarrier frequency, $f_c$ represents the carrier frequency, $B$ is the bandwidth while $\tilde{\theta}_{k}^{s_l}$ represents the squinted beam direction. Furthermore, the normalized array gain (NAG) \cite{dovelos2021channel} corresponding to the $k$-th subcarrier is given by $\Lambda_k = \big\vert \big(\tilde{\mathbf{a}}_{\mathrm{BS}}^{s_l}(\tilde{\theta}_{k}^{s_l},f_k)\big)^H \tilde{\mathbf{a}}_{\mathrm{BS}}^{s_l}(\tilde{\theta}^{s_l},f_c) \big\vert.$ For a narrowband system, when $f_k$ is close to $f_c$; we have $\Lambda_k \approx 1$, which indicates that the NAG is uniform for all the subcarriers. But for a wideband THz system, $f_k$ may significantly deviate from $f_c$ and therefore we have $\tilde{\mathbf{a}}_{\mathrm{BS}}^{s_l}(\tilde{\theta}_k^{s_l},f_k) \neq \tilde{\mathbf{a}}_{\mathrm{BS}}^{s_l}(\tilde{\theta}_k^{s_l},f_c)$. This leads to the beam pointing toward a different direction and gives rise to the \textit{beam-split effect}. Additionally, the band-limited dual-wideband THz channel can be modeled as $\mathbf{H}_u^{s_l}[k] = \mathbf{H}_u^{s_l,\mathrm{LoS}}[k] + \mathbf{H}_u^{s_l,\mathrm{NLoS}}[k]$,
\vspace{-4.5mm}
\begin{align}
    \mathbf{H}_u^{s_l, \mathrm{LoS}}[k] = \sqrt{N_{T,u}N_{\mathrm{BS}}^{s_l}} \alpha(f_k,d)\beta_{\tau}&G_{T,u}G_R^{s_l} \tilde{\mathbf{a}}_{\mathrm{BS}}^{s_l}\big(\tilde{\theta}^{s_l,R},f_k\big) \notag \\ & \tilde{\mathbf{a}}_u^H\big(\tilde{\theta}^{s_l,T},f_k\big), 
\end{align}
\vspace{-8.5mm}
\begin{align}
    \mathbf{H}_u^{s_l,\mathrm{NLoS}}[k] = & \sqrt{\frac{N_{T,u} N_{\mathrm{BS}}^{s_l}}{N_{\mathrm{NLoS}} N_{ray}}} \sum_{q=1}^{N_{\mathrm{NLoS}}} \sum_{j=1}^{N_{ray}} \alpha_{q,j}(f_k, d_{q,j}) \beta_{\tau_{q,j}} \notag \\ & G_{T,u} G_R^{s_l} \tilde{\mathbf{a}}_{\mathrm{BS}}^{s_l}\big(\tilde{\theta}_{q,j}^{s_l,R},f_k\big) \tilde{\mathbf{a}}_u^H\big(\tilde{\theta}^{s_l,T}_q,f_k\big),
\end{align}
where $\beta_{\tau_{q,j}} = \sum_{z=0}^{K-1}p(zT_s - \tau_{q,j})e^{-j\frac{2 \pi k z}{K}}, \forall \, k,z$. The quantities $G_{T,u}$ and $G_R^{s_l}$ represent the transmit and receive antenna gains, while the variables $\tau_{(.)}$ and $ \alpha_{(.)}$ denote the delay and complex-path gain. The quantity $p(.)$ represents the pulse shaping filter, while $T_s$ represents the sampling interval. Furthermore, the quantity $\tilde{\theta}_{(.)}^{s_l,R}$ denotes the AoA, while the variable $\tilde{\theta}_{(.)}^{s_l,T}$ denotes the AoD, where $(.)$ represents the LoS/ NLoS complex path. The calculation of complex-path gain is detailed in our paper \cite{garg2024angularly} and omitted here due to the space constraint. In order to maximize the sum spectral efficiency of \eqref{Spec_eff}, one has to design the optimal TPCs and RCs corresponding to each user. The next section describes the generation of frequency-invariant TPCs/ RCs.
\vspace{-1mm}
\section{True Time Delay (TTD) based two-stage optimization and compensation framework}\label{Dpp}
\subsection{Design of frequency-independent optimal beamformers}
To this end, the singular value decomposition (SVD) of the $u$-th user channel response $\mathbf{H}_u[k]$ can be obtained as
\vspace{-2mm}
\begin{align}
    \mathbf{H}_u[k] = \mathbf{U}[k]\mathbf{\Sigma}[k]\mathbf{V}^H[k],
\end{align}
and the optimal TPC $\mathbf{F}_u^{\mathrm{opt}}[k] \in \mathbb{C}^{N_{T,u} \times N_{s,u}}$ is comprised of the dominant $N_{s,u}$ columns of the right-singular matrix $\mathbf{V}[k]$. However, the SC-FDE-based system is comprised of frequency-independent TPCs, and therefore one can formulate the optimization problem to compute these as
\vspace{-2mm}
\begin{align}
    \big(\mathbf{F}_{\mathrm{RF},u}^{\mathrm{opt}}, \mathbf{F}_{\mathrm{BB},u}^{\mathrm{opt}}\big) \approx \mathop{\mathrm{arg \: min}} \limits_{\mathbf{F}_{\mathrm{RF},u}, \mathbf{F}_{\mathrm{BB},u}} \left\Vert \mathbf{F}_u^{\mathrm{opt}} - \mathbf{F}_{\mathrm{RF},u}\mathbf{F}_{\mathrm{BB},u} \right\Vert_\mathrm{F}^2,\notag
\end{align}
where $\mathbf{F}_u^{\mathrm{opt}} = \frac{1}{K}\sum_{k=0}^{K-1} \mathbf{F}_u^{\mathrm{opt}}[k]$ \cite{srivastava2021fast}. Moreover, it can be readily seen that $\mathcal{C}(\mathbf{F}_u^{\mathrm{opt}}) \subset \mathcal{R}(\mathbf{H}_u[k])$, which further implies that the columns of $\mathbf{F}_{\mathrm{RF},u}$ can be chosen from the dictionary matrix $\mathbf{A}_u(\Theta_u,f_c) \in \mathbb{C}^{N_{T,u} \times G_u}$, where $\Theta_u$ represents the transmit tr's angular grid given by $\Theta_u = \big\{\theta_u: \cos{\theta_u} = \frac{2}{G_u}(u-1)-1, 1 \leq u \leq G_u \big\}$. Here $G_u$ represents the transmit angular grid-size. Therefore, one can further define the transmit dictionary matrix as
\vspace{-2mm}
\begin{equation}
    \begin{aligned}
    \mathbf{A}_u(\Theta_u, f_c) = \big[\tilde{\mathbf{a}}_{T,u}(\theta_{1,u},f_c),\cdots, \tilde{\mathbf{a}}_{T,u}(\theta_{G_u,u},f_c)\big].
\end{aligned}
\end{equation}
The above optimization problem can be solved using the popular simultaneous orthogonal matching pursuit (SOMP) \cite{el2014spatially} technique, where the columns of $\mathbf{F}_{\mathrm{RF},u}^{\mathrm{opt}}$ are chosen from the transmit dictionary matrix. Let $\tilde{\Theta}_u^{Tr}$ represent the set of angles corresponding to the selected columns of $\mathbf{A}_u(\Theta_u,f_c)$. These will be further used for calculating the optimal time delays corresponding to each RF chain in Stage-$2$. Let $\mathbf{H}_{\mathrm{eq}}[k] = \mathbf{H}_{\mathrm{MU}}[k]\tilde{\mathbf{F}}_{\mathrm{RF}}\tilde{\mathbf{F}}_{\mathrm{BB}} \in \mathbb{C}^{N_{\mathrm{BS}} \times N_s}$ represent the equivalent channel response. One can now formulate the optimal MMSE combiner $\mathbf{W}_{\mathrm{opt}}[k] \in \mathbb{C}^{N_{\mathrm{BS}} \times N_s}$ as
\vspace{-2mm}
\begin{align}
    \mathbf{W}_{\mathrm{opt}}[k] = \mathbf{H}_{\mathrm{eq}}[k]\left(\mathbf{H}_{\mathrm{eq}}^H[k]\mathbf{H}_{\mathrm{eq}}[k] + \sigma^2 N_s \mathbf{I}_{N_s}\right)^{-1}.
\end{align}
Additionally, the concatenated optimal RC $\mathbf{W}_{\mathrm{opt}} \in \mathbb{C}^{N_{\mathrm{BS}} \times K N_s}$ and the concatenated baseband RC $\mathbf{W}_{\mathrm{BB}} \in \mathbb{C}^{N_{\mathrm{RF}}^R \times K N_s}$ are correspondingly defined as $ \mathbf{W}_{\mathrm{opt}} = \left[\mathbf{W}_{\mathrm{opt}}[1], \mathbf{W}_{\mathrm{opt}}[2], \cdots, \mathbf{W}_{\mathrm{opt}}[K] \right],
    \mathbf{W}_{\mathrm{BB}} = \left[\mathbf{W}_{\mathrm{BB}}[1], \mathbf{W}_{\mathrm{BB}}[2],\cdots,\mathbf{W}_{\mathrm{BB}}[K]\right].$ Let $\tilde{\mathbf{W}}_{\mathrm{BB}} = \mathbf{A} \mathbf{W}_{\mathrm{BB}} \in \mathbb{C}^{N_{\mathrm{RF}}^R \times K N_s}$. Therefore, one can compute the best approximation of the hybrid RC by formulating it as the optimization problem of
\vspace{-2mm}
\begin{align}
    \big(\mathbf{W}_{\mathrm{RF}}^{\mathrm{opt}}, \tilde{\mathbf{W}}_{\mathrm{BB}}^{\mathrm{opt}} \big) = \mathop{\mathrm{arg\,min}}\limits_{\big(\mathbf{W}_{\mathrm{RF}},\tilde{\mathbf{W}}_{\mathrm{BB}}\big)} \Vert \mathbf{W}_{\mathrm{opt}} - \mathbf{W}_{\mathrm{RF}} \tilde{\mathbf{W}}_{\mathrm{BB}} \Vert_\mathrm{F}^2. \notag
\end{align}
\begin{algorithm}[t]
\DontPrintSemicolon 
\KwIn{$\mathbf{W}_{\mathrm{opt}}, \mathbf{A}_{\mathrm{BS}}(\Theta_b,f_c)$}
\textbf{Initialization:} $\mathbf{P}^{(0)} = \mathbf{W}_{\mathrm{opt}}, \mathbf{W}_{\mathrm{RF}} = [\ ],\tilde{\mathbf{W}}_{\mathrm{RF}} = [\ ]$
 
\For{$s_l = 1: N_{\mathrm{RF}}^R$}
{
{\small $\tilde{\mathbf{A}}_{\mathrm{BS}}(\Theta_b,f_c) = \mathbf{A}_{\mathrm{BS}}(\Theta_b,f_c)\big[:,(1+(s_l-1)G_{\mathrm{BS}}^s : s_lG_{\mathrm{BS}}^s) \big]$}

$\tilde{\mathbf{\Phi}}^{(s_l)} = \tilde{\mathbf{A}}_{\mathrm{BS}}^H(\Theta_b,f_c) \mathbf{P}^{(s_l-1)}$
 
$i^{(s_l)} = \mathop{\mathrm{arg \, max}} \limits_{j=1:G_{\mathrm{BS}}^{s_l}}\big[\mathrm{diag}\big(\tilde{\mathbf{\Phi}}^{(s_l)}\big(\tilde{\mathbf{\Phi}}^{(s_l)}\big)^H\big)\big]$

$\mathbf{W}_{\mathrm{RF}}^{(s_l)} = \big[\mathbf{W}_{\mathrm{RF}}, \tilde{\mathbf{A}}_{\mathrm{BS}}(\Theta_b,f_c)[:,i^{(s_l)}] \big]$

$\mathbf{W}_{\mathrm{BB}}^{(s_l)} = \big[ \big(\mathbf{W}_{\mathrm{RF}}^{(s_l)}\big)^H \mathbf{W}_{\mathrm{RF}}^{(s_l)}\big]^{-1} \big(\mathbf{W}_{\mathrm{RF}}^{(s_l)}\big)^H \mathbf{W}_{\mathrm{opt}}$

$\mathbf{P}^{(s_l)} = \frac{\mathbf{W}_{\mathrm{opt}} - \mathbf{W}_{\mathrm{RF}}^{(s_l)} \mathbf{W}_{\mathrm{BB}}^{(s_l)}}{\left\Vert \mathbf{W}_{\mathrm{opt}} - \mathbf{W}_{\mathrm{RF}}^{(s_l)} \mathbf{W}_{\mathrm{BB}}^{(s_l)} \right\Vert_\mathrm{F}}$

$\tilde{\mathbf{W}}_{\mathrm{RF}} = \big[ \tilde{\mathbf{W}}_{\mathrm{RF}}, \mathbf{A}_{\mathrm{BS}}^{s_l}(\Theta_b, f_c)[:,i^{(s_l)}] \big]$

$(\tilde{\Theta}_b^R)[s_l] = \Theta_b \big(i^{(s_l)}\big)$
}

\textbf{return:~~}{$\tilde{\mathbf{W}}_{\mathrm{RF}}, \tilde{\Theta}_b^R$}
\caption{Spatially sparse algorithm for optimal combiner design}
\label{com_1stage}
\end{algorithm}
As detailed in Section-\ref{parsys}, the receiver possesses a partially connected architecture associated with low-resolution ADCs, which precludes the application of the general SOMP algorithm described in \cite{el2014spatially} to solve the above problem. In this regard, we introduce a spatially sparse algorithm similar to \cite{el2014spatially}, which determines the optimal beam steering angle corresponding to each RF chain. Let $\mathbf{A}_{\mathrm{BS}}^{s_l}(\Theta_b, f_c) \in \mathbb{C}^{N_{\mathrm{BS}}^{s_l} \times G_{\mathrm{BS}}^{s_l}}$ represent the dictionary matrix corresponding to the $s_l$-th subarray, where $\Theta_b$ represents the receive angular grid, given as $\Theta_b = \left\{\theta_b: \cos\theta_b = \frac{2}{G^{s_l}_{\mathrm{BS}}}(b-1)-1, 1 \leq b \leq G^{s_l}_{\mathrm{BS}}\right\}$ and $G_{\mathrm{BS}}^{s_l}$ represents the grid-size corresponding to the subarray. Therefore, the receive dictionary matrix corresponding to the $s_l$-th subarray can be constructed as
\vspace{-2mm}
\begin{align}
    \mathbf{A}_{\mathrm{BS}}^{s_l}\left(\Theta_b, f_c\right) = \big[\tilde{\mathbf{a}}\left(\theta_1,f_c\right), \tilde{\mathbf{a}}\left(\theta_2, f_c\right), \cdots, \tilde{\mathbf{a}}\big(\theta_{G_{\mathrm{BS}}^{s_l}},f_c\big) \big]. \notag
\end{align}
Moreover, we can define the concatenated receive dictionary matrix $\mathbf{A}_{\mathrm{BS}}(\Theta_b,f_c)$ corresponding to all $N_{\mathrm{RF}}$ subarrays as
\vspace{-1mm}
\begin{align}
    \mathbf{A}_{\mathrm{BS}}(\Theta_b,f_c) = \mathrm{blkdiag}\big(\mathbf{A}_{\mathrm{BS}}^{s_1}(\Theta_b,f_c),\cdots,\mathbf{A}_{\mathrm{BS}}^{s_{N_{\mathrm{RF}}^R}}(\Theta_b,f_c)\big).\notag
\end{align}
The quantities $\mathbf{W}_{\mathrm{opt}}$ and $\mathbf{A}_{\mathrm{BS}}(\Theta_b,f_c)$ are fed into the spatially sparse algorithm to obtain the optimal RF RC $\mathbf{W}_{\mathrm{RF}}^{\mathrm{opt}}$, which corresponds to a block-diagonal matrix, where each diagonal block is a column of $\tilde{\mathbf{W}}_{\mathrm{RF}}$, and the corresponding angles $\tilde{\Theta}_b^R$ are determined to calculate the optimal time delays in Stage-$2$. The key steps of this novel procedure are outlined in Algorithm \ref{com_1stage}. Moreover, one can further observe that \textit{only a single column} is selected from each subarray dictionary. This practical consideration significantly reduces computational complexity, particularly for large-scale THz arrays. The next subsection generates the frequency dependent precoders/ combiners in order to compensate the beam-split effect.
\begin{figure}
\centering
\includegraphics[scale=0.25]{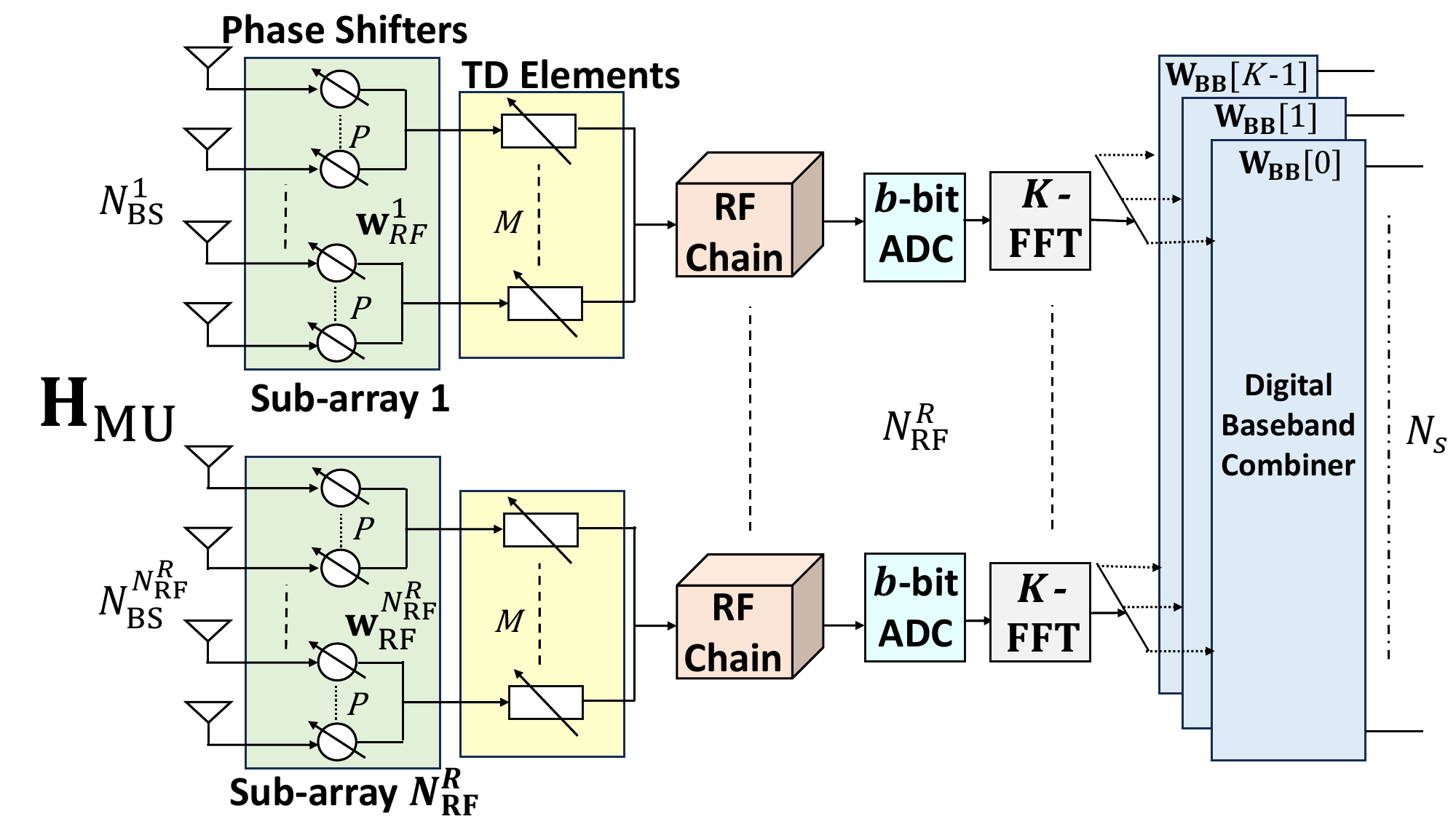}
\vspace{-2mm}
\caption{Subarray-based combining with ADCs and time delay elements}
\label{THz_MIMO}
\vspace{2mm}
\end{figure}
\subsection{Generation of frequency-dependent beamformers}
In this section, the frequency independent optimal precoders and combiners generated in Stage-I are converted to their frequency-dependent counterparts by introducing the optimal time delays. In this regard, let each RF chain be connected to $M$ TD elements that are further connected to $P = \frac{N_{\mathrm{BS}}^{s_l}}{M}$ phase shifters, as shown in Fig. \ref{THz_MIMO}. Let $\theta^{s_l}$ represent the physical target direction, and for simplicity, let $\theta^{s_l} = \sin{\tilde{\theta}^{s_l}}$. Next, to form the frequency-dependent phase shifters, TD elements are inserted at the $s_l$-th RF chain as $
    \tilde{\mathbf{a}}_{1,k} = \tilde{\mathbf{a}}_1 e^{-j2 \pi f_k t^{s_l}_1}, \tilde{\mathbf{a}}_{2,k} = \tilde{\mathbf{a}}_2 e^{-j2 \pi f_k t^{s_l}_2}, \tilde{\mathbf{a}}_{M,k} = \tilde{\mathbf{a}}_M e^{-j 2 \pi f_k t^{s_l}_M}$, where $\tilde{\mathbf{a}}^{s_l} = \left[\tilde{\mathbf{a}}_1^T, \tilde{\mathbf{a}}_2^T,\cdots,\tilde{\mathbf{a}}_M^T \right]^T \in \mathbb{C}^{N_{\mathrm{BS}}^{s_l} \times 1}$ represents the array response vector corresponding to subarray $s_l$ having $M$ sub-vectors consisting of equal numbers of phase elements, where $\tilde{\mathbf{a}}^T_{(.)} \in \mathbb{C}^{P \times 1}$. Therefore, the frequency-dependent phase shifter $\tilde{\mathbf{a}}_{k}^{s_l}$ corresponding to the $s_l$-th sub-array and $k$-th subcarrier is given by
\vspace{-2mm}
\begin{align}
    \tilde{\mathbf{a}}_{k}^{s_l} = \mathrm{blkdiag}\left( \tilde{\mathbf{a}}_1^T, \tilde{\mathbf{a}}_2 ^T,\cdots,\tilde{\mathbf{a}}_M^T\right) e^{-j 2 \pi f_k \mathbf{t}^{s_l}}, \label{general_array}
\end{align}
where $\mathbf{t}^{s_l} = \left[t_{1}^{s_l}, t_{2}^{s_l}, \cdots, t_{M}^{s_l} \right]^T$ represents the TD elements corresponding to the $s_l$-th RF chain. Let $\mathbf{p}_{k}^{s_l} = e^{-j 2 \pi f_k \mathbf{t}^{s_l}} \in \mathbb{C}^{M \times 1}$ represent the frequency-dependent phase shifters realized by TD elements. Moreover $\mathbf{p}_{k}^{s_l}$ shares the same form as the analog beamforming vector $\tilde{\mathbf{a}}^{s_l}$ in order to maintain directivity. Therefore, one can define $\mathbf{p}_{k}^{s_l}$ as
\vspace{-0.5mm}
\begin{align}
    \mathbf{p}_{k}^{s_l} = \big[1, e^{-j \pi \vartheta_{k}^{s_l}}, \cdots, e^{-j \pi (M-1) \vartheta_{k}^{s_l}} \big]^T, \label{freq-delay}
\end{align}
where $\vartheta_{k}^{s_l} \in \left[-1,1\right]$ represents the beamsteering factor for the $k$-th subcarrier corresponding to the $s_l$-th RF chain. Therefore, by adjusting the rotation factor $\vartheta_{k}^{s_l}$, the beams generated by $\tilde{\mathbf{a}}_{k}^{s_l}$ can be aligned with the physical target direction corresponding to all the subcarriers. Moreover, the angle at which the maximum array gain is obtained by solving
\vspace{-1.5mm}
\begin{align}
    \varphi_{\mathrm{opt}} = \mathop{\mathrm{arg \, max}} \limits_\theta \big\vert \tilde{\mathbf{a}}_{\mathrm{BS}}^{s_l}\left(\theta,f_k\right)\tilde{\mathbf{a}}_{\mathrm{BS}}^{s_l}\big(\tilde{\theta}^{s_l},f_c\big) \big\vert.
\end{align}
Therefore, the maximum array gain is obtained, when $\big(\theta^{s_l} - \frac{f_k}{f_c} \varphi_{\mathrm{opt}} \big) P + \vartheta_k^{s_l} = 0$, which reduces to
\vspace{-3mm}
\begin{align}
    \varphi_{\mathrm{opt}} = \frac{\theta^{s_l}}{\frac{f_k}{f_c}} + \frac{\vartheta_{k}^{s_l}}{\frac{f_k}{f_c}P}.
\end{align}
Thus, by setting the optimal direction $\varphi_{\mathrm{opt}}$ identical to the physical direction $\theta^{s_l}$, one can obtain $\vartheta_{k}^{s_l} = \big(\frac{f_k}{f_c}-1\big)P \theta^{s_l}$ which indicates that the array gain due to the beam-split effect can be effectively eliminated using TD elements. Moreover, observe from Equation \eqref{freq-delay}, that the phase difference between the adjacent TTD elements should be equal to the TD vector $\mathbf{t}^{s_l}$ that satisfies the form
\vspace{-2mm}
\begin{align}
    \mathbf{t}^{s_l} = \left[0,n\,T_c,2\,n\,T_c,\cdots,(M-1)\,n\,T_c\right]^T, \label{time_delay}
\end{align}
\hspace{-1mm}where $T_c$ is the period of the carrier frequency and $n$ is the number of periods that should be delayed corresponding to the $s_l$-th RF chain. Therefore, one can also express the frequency-dependent phase shifters as
\begin{algorithm}[t]
\DontPrintSemicolon 
\KwIn{$\mathbf{H}_u[k], \mathbf{F}_{\mathrm{RF},u}, \tilde{\Theta}_u^T$}
 
\For{$l = 1, 2, \cdots, N_{\mathrm{RF},u}^T$}
{
 $\tilde{\mathbf{a}}^{l} = \mathbf{F}_{\mathrm{RF},u}[:,l]; \theta^{l, T} = \tilde{\Theta}^{T}_u[l]$

  \For{$m = 1,2,\ldots,M$}
 {
$\tilde{\mathbf{a}}^{l}_m = \tilde{\mathbf{a}}^{l}[1+(m-1)P:mP] e^{j \pi (M-1) P \theta^{l, T}}$
 
Calculate the time delay $\tilde{t}_m^{l}$ using equation \eqref{final_TD}

$\tilde{\mathbf{a}}_{m,k}^{l} = \tilde{\mathbf{a}}_m^{l} e^{-j 2 \pi f_k \tilde{t}_m^{l}}$
}

$\mathbf{F}_{\mathrm{RF},u}[k](:,l) = \big[(\tilde{\mathbf{a}}_{1,k}^{l})^T, (\tilde{\mathbf{a}}_{2,k}^{l})^T, \cdots,(\tilde{\mathbf{a}}_{M,k}^{l})^T \big]^T$
 
}

$\tilde{\mathbf{H}}_{\mathrm{eq}}[k] = \mathbf{H}_u[k] \mathbf{F}_{\mathrm{RF},u}[k]$

$\tilde{\mathbf{H}}_{\mathrm{eq}}[k] = \mathbf{U}_{\mathrm{eq}}[k] \mathbf{\Sigma}_{\mathrm{eq}}[k] \mathbf{V}_{\mathrm{eq}}^H[k]$

$\mathbf{F}_{\mathrm{BB},u}[k] = \mathbf{V}_{\mathrm{eq},[:,1:N_s^u]}[k]$

\textbf{return:~~}{$\mathbf{F}_{\mathrm{RF},u}[k], \mathbf{F}_{\mathrm{BB},u}[k]$}
\caption{TTD based frequency dependent precoder design}
\label{pre_2stage}
\end{algorithm}
\vspace{-2mm}
\begin{align}
    \mathbf{p}_{k}^{s_l} = \big[1,e^{-j 2 \pi f_k n  T_c},\cdots,e^{-j 2 \pi f_k (M-1) n  T_c} \big]^T. \label{TD}
\end{align}
On comparing Equation \eqref{freq-delay} and \eqref{TD}, one obtains the number of periods as $n = \frac{P \theta_l}{2 \frac{f_k}{f_c}}\big(\frac{f_k}{f_c} - 1 \big)$. It is important to note that the quantity $n$ depends on $P$, $\theta^{s_l}$ and the relative frequency $\frac{f_k}{f_c}$. Hence the proposed beamforming technique is challenging to realize for all the $K$ subcarriers. The expression for $n$ can be simplified as $n = \frac{P \theta^{s_l}}{2} - \frac{P \theta^{s_l}}{2\frac{f_k}{f_c}}$. By substituting $n$ into Equation \eqref{time_delay}, the time delay corresponding to the $s_l$-th RF chain and $m$-th delay element can be expressed as
\begin{align}
       t_{m}^{s_l} = (m-1)nT_c = (m-1)\frac{P \theta^{s_l}}{2}T_c - (m-1) \frac{P \theta^{s_l}}{2 \frac{f_k}{f_c}}T_c, \notag
\end{align}
\normalsize {where the first term $\tilde{t}_{m}^{s_l} = (m-1)\frac{P \theta^{s_l}}{2}T_c, \: \forall \: 1 \leq m \leq M$, is fixed for all the subcarriers, while the second term is variable and can be realised by adding an extra phase shift. Hence, the overall array response vector $\tilde{\mathbf{a}}_{k}^{s_l}$ after incorporating the TD elements in Equation \eqref{general_array} is given by}
{\small \begin{align}
    \tilde{\mathbf{a}}_{k}^{s_l} = \mathrm{blkdiag}\big( \tilde{\mathbf{a}}_1^T,\tilde{\mathbf{a}}_2^T e^{j \pi P \theta^{s_l}}, \cdots, \tilde{\mathbf{a}}_M^T e^{j \pi (M-1) P \theta^{s_l}} \big) e^{-j 2 \pi f_k \check{\mathbf{t}}^{s_l}}, \notag
\end{align}}
\normalsize
\hspace{-2mm}where $\check{\mathbf{t}}^{s_l} = \left[\tilde{t}_{1}^{s_l},\tilde{t}_{2}^{s_l},\cdots,\tilde{t}_{M}^{s_l} \right]$ represents the actual time delay vectors. In general, the time delays $\tilde{t}_m^{s_l}$ are positive and therefore, the generalized expression corresponding to the $m$-th delay element and RF chain $s_l$ can be formulated as
\begin{figure*}
	\centering
	\subfloat[]{\includegraphics[scale=0.315]{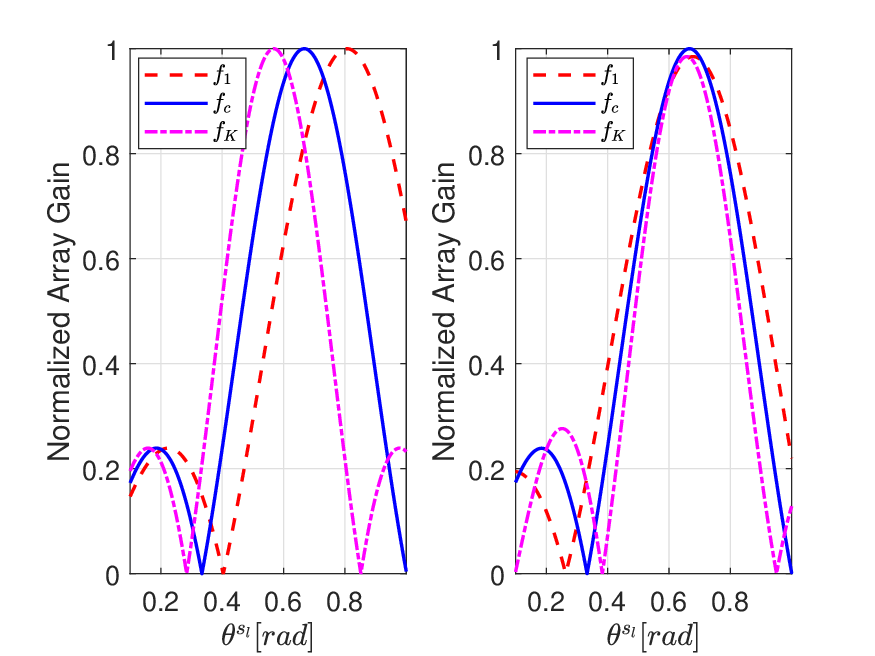}}
	\hfil
	\hspace{-16pt}\subfloat[]{\includegraphics[scale=0.315]{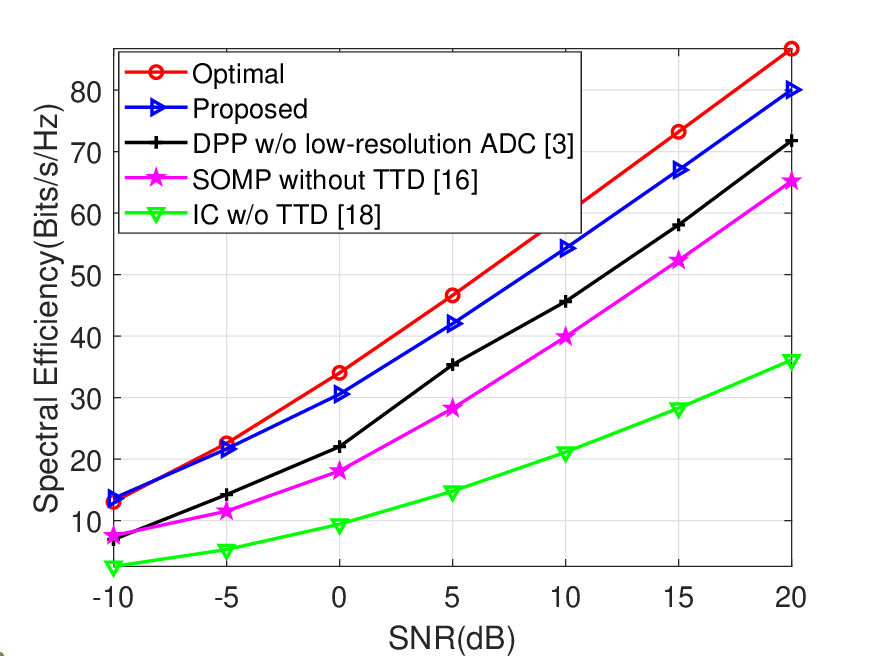}}
 	\hfil
	\hspace{-16pt} \subfloat[]{\includegraphics[scale=0.315]{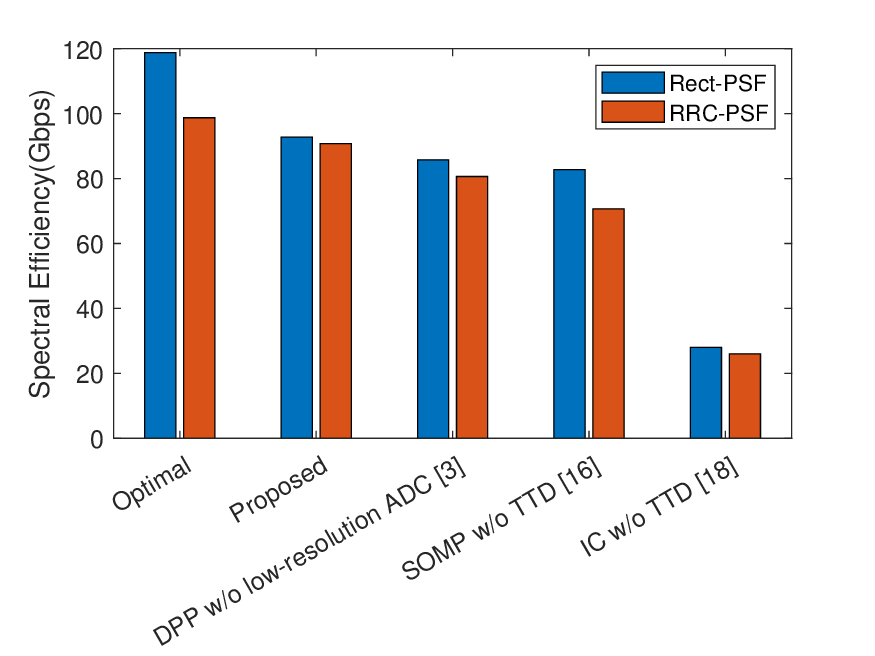}}
 \hspace{-12pt} \subfloat[]
     {\includegraphics[scale=0.315]{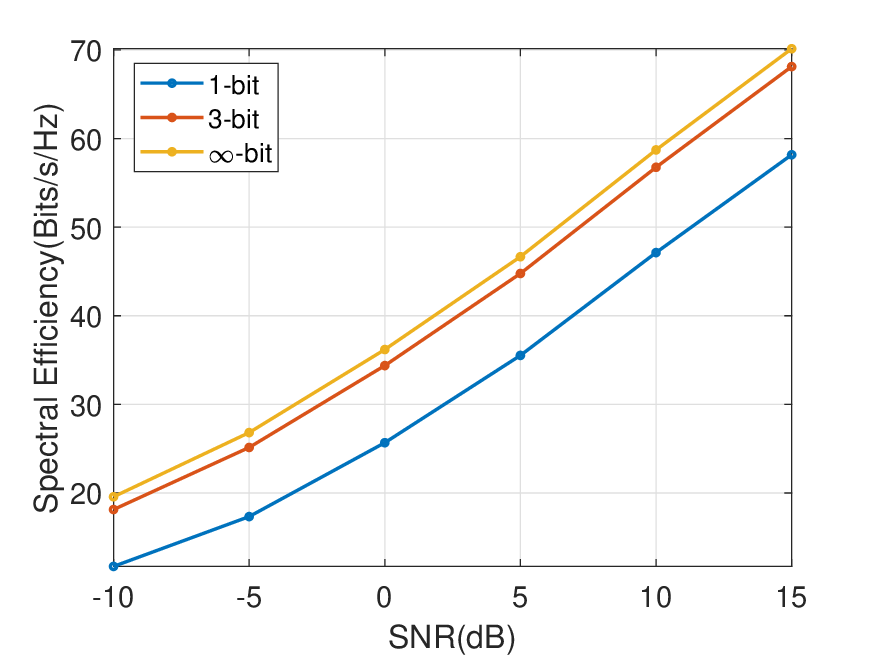}
    }
  \vspace{-1.5mm}
	\caption{$\left(a\right)$ NAG versus physical direction comparison (left) considering traditional precoding (right) for the proposed beamforming technique $ \left(b\right) $ Spectral efficiency comparison versus SNR for the proposed and existing state-of-the-art techniques. Note that, for simulation purposes, we have modified their existing system and channel models to ensure a fair comparison. $ \left(c\right) $ Spectral efficiency comparison for the RRC-PSF based and Rect-PSF based dual-wideband channels with different transceiver schemes. $\left(d\right)$ The spectral efficiency vs. SNR(dB) for different ADC resolutions.}\vspace{-1\baselineskip}
	\label{com_fig}
\end{figure*}
\vspace{-1.5mm}
\begin{align}
    \tilde{t}_m^{s_l} = \scalebox{0.95}{$\begin{cases}
        (m-1)\frac{P \theta^{s_l}}{2}T_c, \quad\quad\quad\quad\quad\quad\quad\quad\:\:\: \theta^{s_l}\geq 0,  \\
    (M-1)\left \vert \frac{P \theta^{s_l}}{2} \right \vert T_c + (m-1) \frac{P \theta^{s_l}}{2} T_c, \: \theta^{s_l} < 0, \label{final_TD}
    \end{cases}$} 
\end{align}
where the second term arises due to the periodicity in the delay elements. Algorithm-\ref{pre_2stage} summaries the design procedure. It is interesting to note that the combined effect of the frequency-independent phase shifters and TTD element at the transmitter creates \textit{frequency-dependent} TPCs. Similar to Algorithm-\ref{pre_2stage}, the TTD based frequency-dependent RCs are designed and omitted due to space constraints. One should note that the partially connected architecture offers potential advantages in the form of a simple design and fabrication, and particularly well-suited for large THz applications.
\vspace{-2mm}
\section{Simulation Results} \label{simulation}
\vspace{-1mm}
We consider the practical scenario of $U = 4$ users each employing $N_{T,u} = 4$ TAs having $N_{\mathrm{RF},u}^T = 2$ RF chains. The BS is equipped with $N_{\mathrm{BS}} = 96$ RAs along with $N_{\mathrm{RF}}^R = 16$ RF chains, where each subarray consists of $N_{\mathrm{BS}}^{s_l} = \frac{N_{\mathrm{BS}}}{N_{\mathrm{RF}}^R} = 6$ antennas with $M = 2$ time delay elements per subarray. The number of data streams corresponding to each user is $N_{s,u} = 2$, while the total number of data streams at the BS is $N_s = U N_{s,u} = 8$. The carrier frequency is $f_c = 1 \, \mathrm{THz}$ with antenna spacing of $\approx \frac{\lambda}{2}$ and the communication distance is $d = 15 \, \mathrm{m}$. We consider a single LoS path and $3$ NLoS paths with $N_{ray} = 1$ diffused ray and $L = 4$ channel taps. We opt for $K = 128$ subcarriers with a total bandwidth of $B = 10 \, \mathrm{GHz}$. The transmit dictionary includes $G_u = 8$ atoms, while the receive dictionary at each subarray includes $G_{\mathrm{BS}}^{s_l} = 12$ atoms. The transmit gain combined for all the users is $\sum_{u=1}^U G_{T,u} = 28 \, \mathrm{dBi}$, while the receiver gain is $G_R^{s_l} = 28 \, \mathrm{dBi}$. Note that the choice of transmit and receive antenna gain is practically valid, since
 the authors in \cite{zhu2020compact} achieve a beamforming gain of nearly 20 dBi for a single-antenna in the $1$ THz range. The SNR is defined as $\mathrm{SNR} = 10 \mathrm{log}_{10}(\frac{1}{\sigma^2})$ where $\sigma^2$ denotes the noise variance. Fig. \ref{com_fig} (a) compares the NAG performance vs. the physical direction. Observe that prior to applying the proposed beamforming technique, the beams are separated from the central frequency $f_c$ which can be attributed to the beam-split effect. By contrasts upon applying the proposed beamforming technique, one can observe that the minimum and maximum subcarrier frequencies are aligned with the physical target direction $\theta^{s_l}$. Fig. \ref{com_fig} (b) demonstrates the significant improvement in the achievable rate compared to the existing state-of-the-art beamforming techniques versus the SNR power. The proposed technique exhibits a performance closer to the capacity limit, due to its higher gains by effectively eliminating the beam split effect of the THz channel. Furthermore, we obtain a relative performance improvement of approximately $13\%$ over the state of the art \cite{dai2022delay}. Fig. \ref{com_fig} (c) compares the attainable rate (in Gbps) of the proposed beamforming technique using the existing DPP \cite{dai2022delay}, SOMP \cite{el2014spatially} and interference cancellation (IC) technique \cite{li2016robust} for the RRC-PSF based dual-wideband channel and the Rect-PSF based dual-wideband channel \cite{dovelos2021channel}. The latter channel exhibits spectral leakage, leading to increased interference levels between neighboring channels. By contrast, channels utilizing RRC-PSF are susceptible to edge attenuation, which may degrade signal recovery at the receiver. Consequently, there exists a clear trade-off between the RRC-PSF and Rect-PSF based dual-wideband channel formulations. Furthermore, the improved performance over the other techniques clearly demonstrates a near perfect elimination of the beam-split effect. Note that all the results are obtained for a $3$-bit ADC. Fig. \ref{com_fig}(d) illustrates the efficacy of the proposed beamforming strategy using low-resolution ADCs. It is observed that a $3$-bit ADC yields a performance very closes to that of an ideal $\infty$-bit quantizer with a marginal loss of only $2$\% at high SNR, making it extremely well-suited for practical systems.
\vspace{-2mm}
\section{Conclusion} \label{conclusion}
\vspace{-1mm}
A novel beamforming technique was conceived for SC-FDE-based MU THz dual-wideband systems, relying on a partially-connected architecture at the BS using low-resolution ADCs. Stage-$1$ computes the optimal precoders/ combiners, while Stage-$2$ transforms the optimal precoders/ combiners to their frequency-dependent equivalents. The proposed beamformer design efficiently mitigates the beam-split effect using very few TTD lines, thus leading to low costs and complexity. Potential future work in this area could include hybrid beamformer design relying on realistic channel estimation.
\vspace{-0.5\baselineskip}
\bibliographystyle{IEEEtran}
\bibliography{References}
\end{document}